\documentclass[a4paper,aps,pre,singlecolumn,showpacs,superscriptaddress,groupedaddress]{revtex4-1}

\usepackage{amsmath}
\usepackage{amssymb}
\usepackage{graphicx}
\usepackage{color}
\usepackage{hyperref}
\usepackage{multirow}
\usepackage{subcaption}

\begin{document}
\title{Large-scale flow in a cubic Rayleigh-B\'{e}nard cell: Long-term turbulence statistics and Markovianity of macrostate transitions}
\author{Priyanka Maity$^{1}$}
\email{priyanka.maity@tu-ilmenau.de}
\affiliation{$^{1}$Institute of Thermodynamics and Fluid Mechanics, Technische Universit\"at Ilmenau, Postfach 100565, D-98684 Ilmenau, Germany}
\author{P\'eter Koltai$^{2}$}
\email{peter.koltai@fu-berlin.de}
\affiliation{Department of Mathematics, Freie Universit\"at Berlin, Arnimallee 6, D-14195 Berlin, Germany}
\author{J\"org Schumacher$^{1,3}$}
\email{joerg.schumacher@tu-ilmenau.de}
\affiliation{$^{1}$Institute of Thermodynamics and Fluid Mechanics, Technische Universit\"at Ilmenau, Postfach 100565, D-98684 Ilmenau, Germany\\
$^{3}$Tandon School of Engineering, New York University, New York, NY 11201, USA} 

\begin{abstract}
{We investigate the large-scale circulation (LSC) in a turbulent Rayleigh-B\'enard convection flow in a cubic closed convection cell by means of direct numerical simulations at a Rayleigh number $Ra=10^6$. The numerical studies are conducted for single flow trajectories up to $10^5$ convective free-fall times to obtain a sufficient sampling of the four discrete LSC states, which can be summarized to one macrostate, and the two crossover configurations which are taken by the flow in between for short periods. We find that large-scale dynamics depends strongly on the Prandtl number $Pr$ of the fluid which has values of 0.1, 0.7, and 10. Alternatively, we run an ensemble of 3600 short-term direct numerical simulations to study the transition probabilities between the discrete LSC states. This second approach is also used to probe the Markov property of the dynamics. Our ensemble analysis gave strong indication of Markovianity of the transition process from one LSC state to another, even though the data are still accompanied by considerable noise. It is based on the eigenvalue spectrum of the transition probability matrix, further on the distribution of persistence times and the joint distribution of two successive macrostate persistence times.}
\end{abstract}


\maketitle
Turbulent convection is a classic example of dynamical system driven out of equilibrium which is omnipresent in nature; astrophysical~\cite{spiegel_1971}, atmospheric, and oceanic convection~\cite{marshall_schott_1999} are the primary contributing factors for observed phenomena in planets and stars. Contrary to the notion of presence of multiple scales and vigorous mixing due to turbulence, natural turbulent convection frequently exhibits formation of large scale structures and patterns, such as in clouds~\cite{markson_1975} and supergranules in Sun~\cite{hathaway_2012,schumacher_2020}. On a physically attainable scale, such natural turbulent convection can be investigated using the simplified Rayleigh--B\'enard convective (RBC) system, where a layer of fluid is confined between two plates with a thermal gradient~\cite{chandrashekhar_1961,ahlers_2009,chilla2012}. Thermal convection originates due to density differences in the fluid as a result of the constant thermal gradient across the fluid layer, represented by the dimensionless Rayleigh number, which is given by 
\begin{equation}
Ra = \frac{\alpha g\, \delta T d^3}{\nu \kappa}\,,
\end{equation}
where $\alpha$ represents the isobaric thermal expansion coefficient, $g$ the acceleration due to gravity, $\delta T=T_{\rm bottom}-T_{\rm top}$ the temperature difference maintained along the fluid layer of thickness $d$, and $\nu$ and $\kappa$ being the kinematic viscosity and thermal diffusivity of the fluid, respectively. Convection begins when the temperature difference between the plates exceeds the critical value, $\delta T >\delta T_c$, which corresponds to the critical Rayleigh number of $Ra_c=1708$ when the impermeable plates satisfy the no-slip boundary condition for the velocity~\cite{chandrashekhar_1961}. The onset is generally characterized by laminar structures. Subsequent increase of $Ra$ leads to a transition from laminar to turbulent behaviour, which is characterized by thinning of boundary layers and generation of spatially extended coherent structures which are now denoted as superstructures of convection \cite{pandey_2018}. 

Turbulent RBC in confined geometries, such as cubic or cylindrical cells, also leads to an alignment of the rising and falling thermal plumes to form a coherent structure known as the large scale circulation or the mean wind of thermal convection \cite{niemela_2001, sreenivasan_2002, parodi_2004, puthenveettil_2005, brown_2006, xi_2007, Verma_book_2018, mishra_2011, verma_2015, mannatil_2017, kumar_2018}. The alignment of the plumes into an LSC and the increasing confinement towards a space direction can significantly influence the heat transport in the system~\cite{bao_2015, wagner_2013, chong_2015}. In such cases the aspect ratio, which denotes the ratio of the horizontal extension (diameter or side length) to the height $H$, is decreased. A further perspective on this subject is taken in refs. \cite{daya_2001,hartmann_2021} and asked how the specific geometry at same aspect ratio alters the heat transfer. 

In cubic closed cells, the LSC cannot drift azimuthally, but appears in a finite number of discrete macroscopic flow states~\cite{foroozani_2014, bai_2016, foroozani_2017, giannakis_2018, vasilev_2019}. In detail, the LSC appears preferentially in the form of circulation rolls along the two diagonals and only an additional weak side wall heating would align the LSC preferentially with the side walls \cite{teimurazov_2021}. This results in four states if one  takes two orientations along each diagonal into account. Fast switches between these states proceed via four transient LSC states that are aligned with the side faces (or edges) of the cell. {While the configurations parallel to the diagonals are considered as long-lived LSC states of the turbulent convection flow, the ones along the side faces of the closed cube are stable, short-lived convection states. Similar transitions between LSC states have also been analysed energetically by Fourier-mode spectral expansions in other convection flow configurations~\cite{mishra_2011,verma_2015,mannatil_2017} where they were termed as ``cessation led flow-reversals". In the system under consideration rather re-orientations between different long-lived flow states (which we will discuss in detail) than real reversals were observed. The switching between different macroscopic flow states can also be detected in other fluid flows, e.g. in mixing processes in T-shaped mixers \cite{schikarski_2019}.

One attempt to study the LSC dynamics has been demonstrated in Giannakis \textit{et al} ~\cite{giannakis_2018} by a data-driven analysis of the eigenfunctions of the linear Koopman operator that describes the unitary time evolution of observables of the dynamical system rather than the nonlinear evolution of the states \cite{mezic_2013,williams_2015}. The leading eigenvectors could be assigned to the 4 stable LSC states, the subsequent eigenstates were related to secondary flow structures in form of corner vortices that drive the system from one LSC state to another. However, their simulation time was still too short to obtain a sufficient number of LSC switches; the computation was run for $10^4$ free-fall times, the characteristic convective time unit of the RBC flow. A detailed analysis of these re-orientations sets the motivation of the present work. 

The question that consequently arises is whether the re-orientation process of the LSC is indeed random and all discrete LSC states appear with the same probability? A further point is how this dynamics depends on the dimensionless Prandtl number, which is given by
\begin{equation}
    Pr=\frac{\nu}{\kappa}\,,
\end{equation}
and quantifies the ratio of viscous to temperature diffusion in the fluid.

In this work, we want to study these points by three-dimensional direct numerical simulations in two different ways. First, we follow a single long-term trajectory through the phase space of the convection flow for $10^5$ free-fall times and determine the transitions between the different LSC states. This analysis is conducted for three different flow cases at a Rayleigh number $Ra=10^6$ and Prandtl numbers $Pr=0.1, 0.7$ and 10. We demonstrate a strong dependence of the large-scale flow behavior on $Pr$ at the fixed $Ra$. Secondly,  in the case of $Pr=0.7$ an alternative approach to the LSC dynamics is presented which we term an {\em ensemble simulation}. Therefore, we take a coarse grid long-term trajectory that advances through phase space and start short-term ensemble simulations runs at full resolution from different outputs along the coarse grid run. The assumption in this approach (which cannot be proven) is that the coarse-grid trajectory ``shadows" the true system evolution sufficiently well \cite{sauer_1991}. In the present study, we can actually take the fully resolved trajectory from the first analysis part, in studies with higher Rayleigh numbers this would not be possible. We then test if a Markov State Model (MSM) is able to describe the transitions between different LSC states~\cite{pande_2010, husic_2018}. A time-discrete MSM describes a hopping process where the probability to get into a future state depends on the current state only, see \cite{SchSa13,prinz_2011} for theory, \cite{BoPaNo14} for applications, and~\cite{SchEtAl15,MSMBuilder17} for software implementations. This implies that we have to determine the transition probabilities among the different LSC states which will be described more specifically further below.  Two different models are compared, one with 6 and one with 3 LSC states for which the 4 diagonal states are considered as one macrostate.
The study also requires to determine the correct sampling time along the individual short-term trajectories that form an ensemble. 

Application of MSMs to stochastic system such as molecular kinetics~\cite{prinz_2011} and protein folding~\cite{husic_2018} are known to show promising results, and recently it has been applied to RBC experimental data as well~\cite{KoWe18}. Markovianity of the large-scale dynamics of the turbulent convection flow at hand would imply that the switching between different large-scale configurations is practically memoryless. \emph{If such a property holds, the large-scale flow behaviour and thus the high-dimensional phase space of the convection flow can be sampled more effectively without running a fully resolved long-term simulation.
}

The manuscript is organized as follows. In section 2, we discuss in brief the numerical simulation model. Section 3 analyses the large-scale flow along the individual long trajectory. In section 4, we explain the details of the MSM and probe the Markov property for the present system. The last section contains a summary and a brief outlook. 

\begin{center}
\begin{figure*}[!h]
\centering\includegraphics[width=0.9\textwidth]{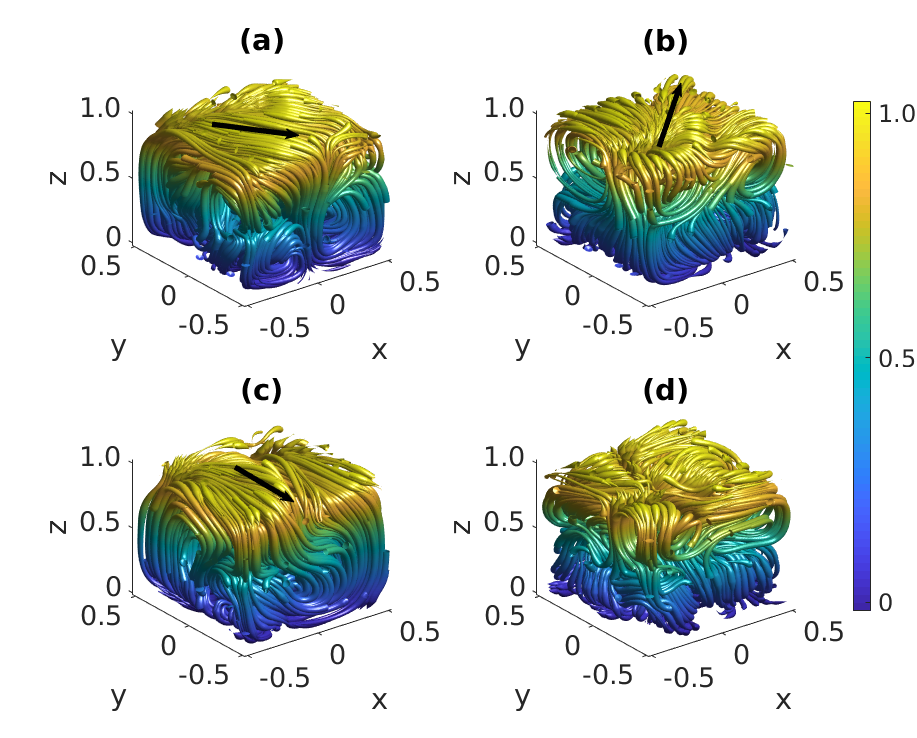}
\caption{Instantaneous velocity streamtubes representing different configurations of the large-scale circulation. (a) Long-lived large-scale circulation (LL-LSC) state S$_{3\pi/4}$ aligned along a diagonal. (b) LL-LSCstate S$_{\pi/4}$ along the other diagonal. (c) Short-lived large-scale circulation (SL-LSC) state S$_{n\pi/2}$ which is aligned along one pair of opposite side faces. (d) Decoherent or Null state S$_0$ without any well-defined large scale circulation. All data are for a fluid with $Pr = 0.7$. The streamtubes are colored with respect to the height $0\le z\le1$, as shown in the legend.}
\label{streamtube}
\end{figure*}
\end{center}

\section{Numerical model}
In the Rayleigh--B\'{e}nard model, the dimensionless equations of motion of an incompressible fluid undergoing thermal convection are given by
\begin{align}
\partial_t {\bf {u}} + (\bf {u \cdot \nabla}) {\bf u} &= -{\bf \nabla} p + \sqrt{\frac {Pr}{Ra}} \nabla^2 {\bf u} +  T \hat{z}\,,\label{NS1} \\
\partial_t T + (\bf {u \cdot \nabla}) T &= {\frac{1}{\sqrt{Ra Pr}}} \nabla^2 T\,. \label{temp1}\\
{\bf \nabla \cdot u} &= 0\,.\label{cont1}
\end{align}
\noindent where ${\bf u}(x,y,z,t) \equiv (u_x(x,y,z,t), u_y(x,y,z,t), u_z(x,y,z,t))$ is the velocity vector field of the fluid, $T = T(x,y,z,t)$ denotes the scalar temperature field, and $p \equiv p(x,y,z,t)$ is the scalar pressure field. We employ the Boussinesq approximation, whereby the density of the fluid $\rho_0$ is assumed to be constant except in the buoyancy term that leads to the last term on the right hand side in \eqref{NS1}. The equations are made dimensionless by scaling the length scales by the thickness of the fluid $d$, velocity by the free-fall velocity $U_f = \sqrt{\alpha g \delta T d}$, and time by the free-fall time scale $t_f = \sqrt{d/(\alpha g \delta T)}$. The Rayleigh number $Ra$ and the  Prandtl number $Pr$ are two control parameters of the system. The aspect ratio of the cube is one. In table \ref{tab1}, we summarize some important statistical quantities of the three simulation runs.   
\begin{table}
\begin{tabular}{lcccccc} 
Run & $Ra$ & $Pr$ & $Nu$ & $Re$ & $u_{\rm rms}$ & $t_{\rm total}/t_f$ \cr 
\hline
1 & $10^6$  & 0.1 & 6.08 & 1550.5 & 0.49 & $10^5$ \cr
2 & $10^6$  & 0.7 & 6.73 & 437.5 & 0.37 & $10^5$ \cr
3 & $10^6$  & 10.0 & 6.92 & 26.3 & 0.08 & $5\times 10^4$ \cr
\end{tabular}
\caption{\label{tab1} Parameters of the simulations. These are the Rayleigh and Prandtl numbers, followed by the Nusselt number $Nu$, the Reynolds number $Re$, the root mean square velocity $u_{\rm rms}$, and the total integration time in units of the free-fall time $t_f$.} 
\end{table}

We performed direct numerical simulations (DNS) of the eqns.~\eqref{NS1}--\eqref{cont1} for three different fluids corresponding to Prandtl numbers of $Pr = 0.1, 0.7$, and $10$ and sustain a Rayleigh number to be $Ra = 10^6$ in each case. Consequently, the effective Reynolds number and the Kolmogorov scales change in each case, allowing us to study the effect of Reynolds number on re-orientations between the large scale circulations. The Reynolds number measures the turbulent momentum transfer in RBC and is given by 
\begin{equation}
    Re=\sqrt{\frac{Ra}{Pr}}\,u_{\rm rms} \quad \mbox{with} \quad u_{\rm rms}=\sqrt{\langle u_x^2+u_y^2+u_z^2\rangle_{V,t}}\,, 
\end{equation}
where $\langle\cdot\rangle_{V,t}$ denotes a combined volume and time average. The variation of $Pr$ alters also the magnitude of the Nusselt number $Nu$, the dimensionless measure of the turbulent heat transfer which is given by
\begin{equation}
    Nu=1+\sqrt{Ra Pr}\langle u_z T\rangle_{V,t}\,,
\end{equation}
see again Table \ref{tab1}.

The simulations were performed using an open source code nek5000 based on a spectral element method \cite{fischer_1997,scheel_2013}. Our simulation volume is a closed cubic container. The system is uniformly heated from below at $z=0$ and cooled from at $z=1$, i.e., Dirichlet conditions for the temperature field apply. We assume no-slip boundary conditions, which translate to $u_x =u_y =u_z = 0$ at all  boundaries. We also assume insulating boundaries for temperature field at the four side faces, i.e., ${\bf n}\cdot \nabla T=0$. We took 16 spectral elements along each direction for simulations and the order of the Lagrangian interpolation polynomials along each space direction and on each spectral element is 5. The vertical profiles of the mean kinetic energy dissipation rate were analyzed to verify that this spectral resolution is sufficient for our purpose. For the LSC flow analysis, we interpolate all vector and scalar fields spectrally onto a uniform mesh. 

In agreement with previous studies~\cite{foroozani_2017,giannakis_2018}, we identify four  long-lived LSC (LL-LSC) states along the diagonals. There are denoted as S$_{\pi/4}$, S$_{3\pi/4}$, S$_{5\pi/4}$, and S$_{7\pi/4}$. Once the flow gets into one of these long-lived states, it stays there for a considerable amount of time before re-orienting into another configuration. As already mentioned in the introduction, these 4 diagonal states can be summarized to one diagonal state for symmetry reasons. We will come back to this point later in the text and proceed for now with 4 diagonal states. The re-orientations between the 4 long-lived large scale circulations transition via the short-lived large scale structures (SL-LSC) aligned along the edges of the cube. These four individual states are summarized to a fifth LSC configuration which is denoted as S$_{n\pi/2}$ with $n=0, 1, 2, 3$. In addition to the stable and unstable LSC states, we identified a sixth state which does not belong to any of the previous states. In this state, termed as the {\em decoherent state} or {\em Null state} S$_0$, the turbulent system does not have any distinct large scale circulation. Figure~\ref{streamtube} provides a visualization of typical stable and unstable large scale circulations for a Prandtl number of 0.7. Table \ref{tab2} summarizes the 6 different LSC states. The identification of the specific states is described in the subsequent section. 

\begin{table}
\begin{tabular}{lcc} 
LSC state & Orientation angle $\theta$& Coherence/Lifetime  \cr 
\hline
S$_{\pi/4}$ & $\pi/4$  & {Coherent/Long} \cr
S$_{3\pi/4}$ & $3\pi/4$  & {Coherent/Long}  \cr
S$_{5\pi/4}$ & $5\pi/4$  & {Coherent/Long}  \cr
S$_{7\pi/4}$ & $7\pi/4$  & {Coherent/Long}  \cr
S$_{n\pi/2}$ & $n\pi/2$ & {Coherent/Short} \cr
S$_0$ & -- & {Decoherent/Short} \cr
\end{tabular}
\caption{\label{tab2} Six large-scale flow states in the cubical convection cell. The orientation angle $\theta$ is given in figure 2.} 
\end{table}
\begin{figure*}[!ht]
\centering\includegraphics[width=0.9\textwidth]{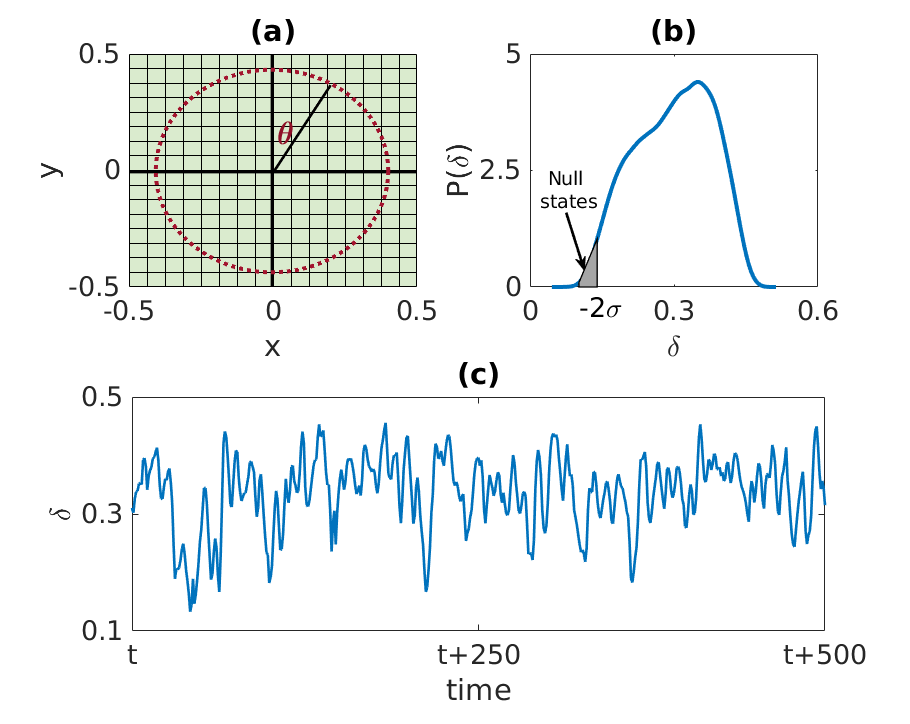}
\caption{(a) Schematic representation of the interpolated grid on the circle from data of the uniform rectangular grid at the mid plane $z=0.5$. The rectangular and circular grids are superimposed on each other for clarity in comprehension. The angle of the LSC ($\theta$) is measured with respect to the origin in clockwise direction. (b) Probability density function (PDF) of the fraction of (vertical) kinetic energy $\delta$ (see eq. \eqref{delta}) carried by the largest Fourier mode. The shaded portion represents the region in which the events were considered as Null states. (c) Shows the temporal variation of $\delta$ over a short window.}
\label{geometry}
\end{figure*}

\begin{figure*}[!ht]
\centering\includegraphics[width=0.45\textwidth]{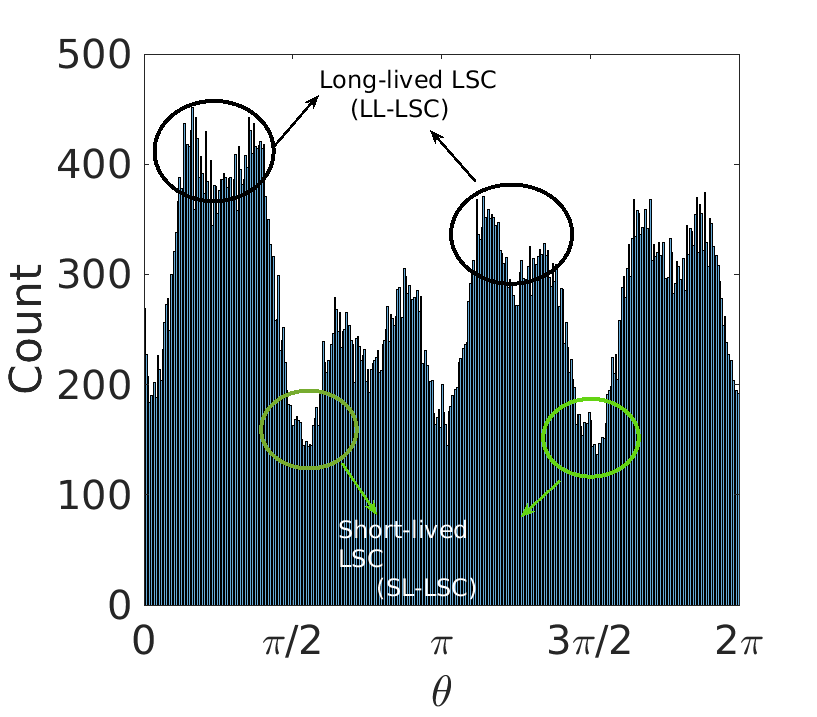} 
\hfill
\centering\includegraphics[width=0.45\textwidth]{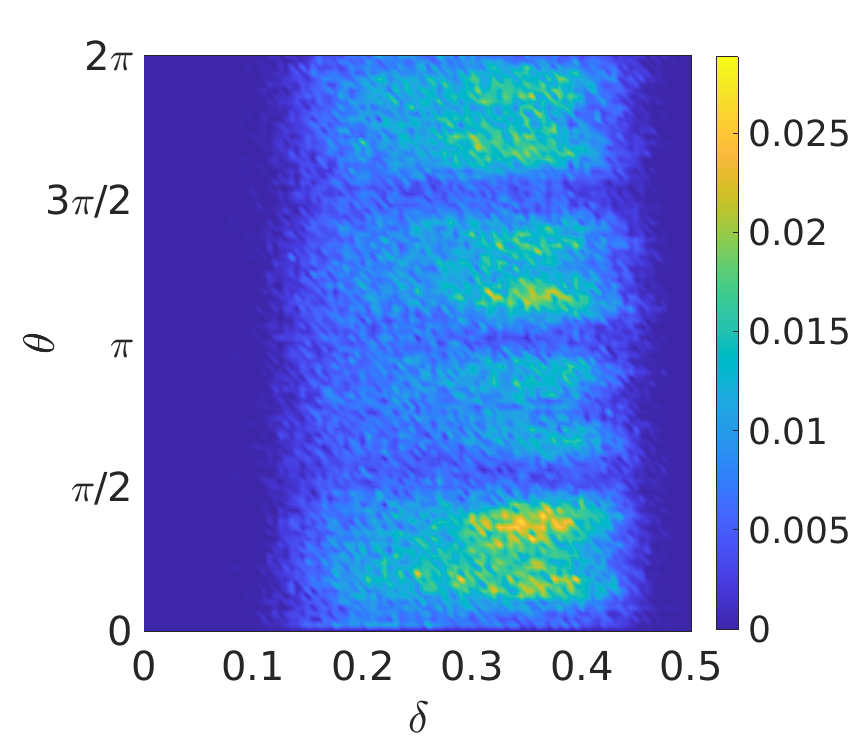}
\caption{ (a) Histogram of orientation angle ($\theta$) of the LSCs as obtained from the long-term trajectory data. (b) Heat map of the joint probability density function of LSC states with respect to the orientation angle ($\theta$) and the fraction of vertical kinetic energy which is contained in the largest Fourier mode ($\delta$).}
\label{macrostates}
\end{figure*}


\section{Large-scale circulation states along single long-term trajectory}\label{LTS}

We began our simulations from a random initial condition and waited for the flow to become fully turbulent. Once the system reached a steady state, we performed the simulations for another $10^5$ free fall times and output the data at each free fall time (in cases $Pr=0.1, 0.7$). For the purpose of measuring the orientation angles of the LSC, we use the vertical velocity component in the mid-plane $z=0.5$ as sketched in figure~\ref{geometry}(a) and done for example in refs.~\cite{foroozani_2017,giannakis_2018}. Thereafter, we interpolated the data from the uniform grid to a circle with fixed radius of $r=0.45$ and an angle $\theta$ (measured with respect to the y-axis in a clockwise manner) varying by $5^{\circ}$ for each subsequent grid point. Thereafter, we obtained the discrete 1d Fourier transform of the vertical velocity component on the circle. 

In presence of a distinct LSC structure in the cubic cell, the largest Fourier mode will possess considerable amount of kinetic energy  and the corresponding phase will give a measure of the angle of orientation ($\theta$) of the LSC. Figure~\ref{geometry}(c) shows the ratio $\delta$ of the vertical kinetic energy carried by the largest Fourier mode to that of the vertical kinetic energy of the whole system (vertical kinetic energy means kinetic energy with respect to $u_z$ only) as a function of time. It is given by 
\begin{equation}
    \delta(t)=\frac{\max_{k_{\theta}} |\hat{u}_z(k_{\theta},t)|^2}{\sum_{k_{\theta}} |\hat{u}_z(k_{\theta},t)|^2}\,.
    \label{delta}
\end{equation}
The majority of the time, the system will be either in one of the four  long-lived LSC states or in the short-lived LSC states S$_{n\pi/2}$. When there is no distinct LSC (Null state), the ratio $\delta$ will be minuscule. This low value of $\delta$ in the Null state S$_0$ is obviously expected as in absence of any LSC structure, the kinetic energy is almost equally distributed among the Fourier modes. These Null states are extremely rare and hence to identify them, we first calculate the probability distribution function (PDF) of the ratio $\delta$ which is shown in figure~\ref{geometry}(b). We then calculated the standard deviation $\sigma$ of the PDF of $\delta$. All the events which fall below  $\mu-2\sigma$ (with $\mu$ being the mean of the PDF) and thus have extremely low values of $\delta$ are identified as the Null states. It is noted that all events larger than $\mu + 2\sigma$  still fall into the category of distinct LSC structures. A visualization of the flow structure in a typical Null state is shown in figure~\ref{streamtube}(d).

As mentioned in the previous sections, we performed the single trajectory analysis in case of fluids with Prandtl numbers of $Pr = 0.1, 0.7$ for $10^5$ free fall time units. We assigned an orientation angle $\theta$ to the LSC structure at each time frame obtained at every free fall time. The classification is conditioned to both, the values of $\theta$ and $\delta$.  For a comprehensible understanding of the prevalence of different states, we plot the histogram of angle of orientation ($\theta$) of the LSC structures in figure~\ref{macrostates}(a) obtained from the single long-term trajectory run. The histogram clearly shows four bi-variate peaks in the vicinity of angles corresponding to $(2n+1)\pi/4$, with $n = 0,1,2,3$. The bi-variate nature of the peaks can be attributed to the fact that corner vortices try to destabilize the LSC structure parallel to the diagonals permanently. Only when the fluctuations are large enough a transition proceeds. Hence, instead of a pronounced LSC structure along the diagonal, we detected oscillations in the angle of orientation with bi-variate peaks. 

Nevertheless, it is clear from the histogram that LSC structures along the diagonals are the most prevalent states and hence can be termed as the ``long-lived LSC (LL-LSC)" states. The minima of the histogram of $\theta$ occurs in the vicinity of $n\pi/2$ with $n=0,1,2,3$, representing the ``short-lived LSC" (SL-LSC) aligned along the edges. A complementary picture of the different macrostates as a function of both, $\theta$ and $\delta$, is given in figure~\ref{macrostates}(b). This panel shows the heat map of joint probability density function (JPDF). The four distinct bands for values $\delta \gtrsim 0.1$ along $\theta \approx (2n+1)\pi/4$ are visible representing the most prevalent LL-LSC structures aligned along the diagonals. The subordinate SL-LSC states are represented by the thin dark bands appearing between the LL-LSC state for $\delta \gtrsim 0.1$ and $\theta \approx n\pi/2$. The rare Null state occurs for $\delta \leq 0.1$. Based on these inferences, we suggest to classify the system into six-basic LSC states.}

\begin{figure*}[!ht]
\centering\includegraphics[width=0.9\textwidth]{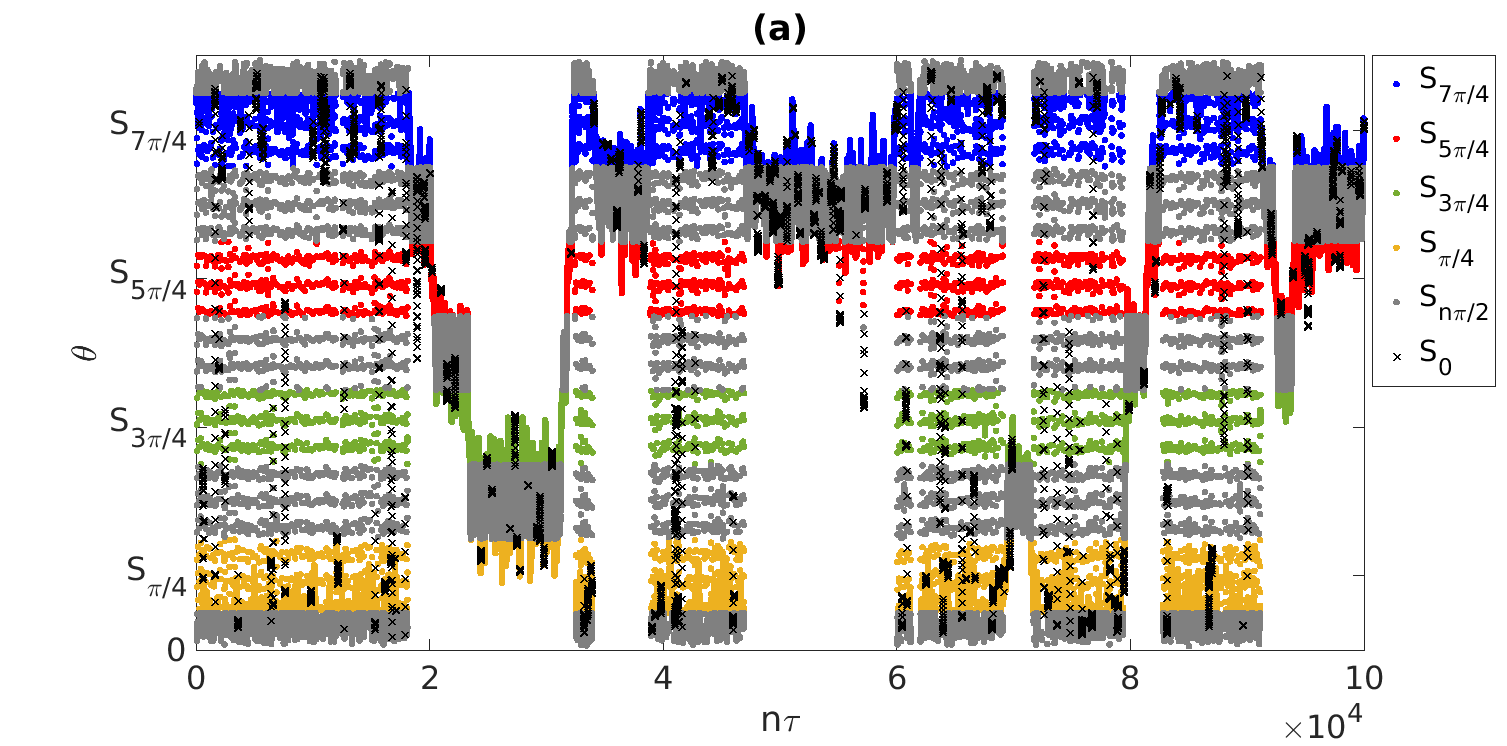}\\
\centering\includegraphics[width=0.9\textwidth]{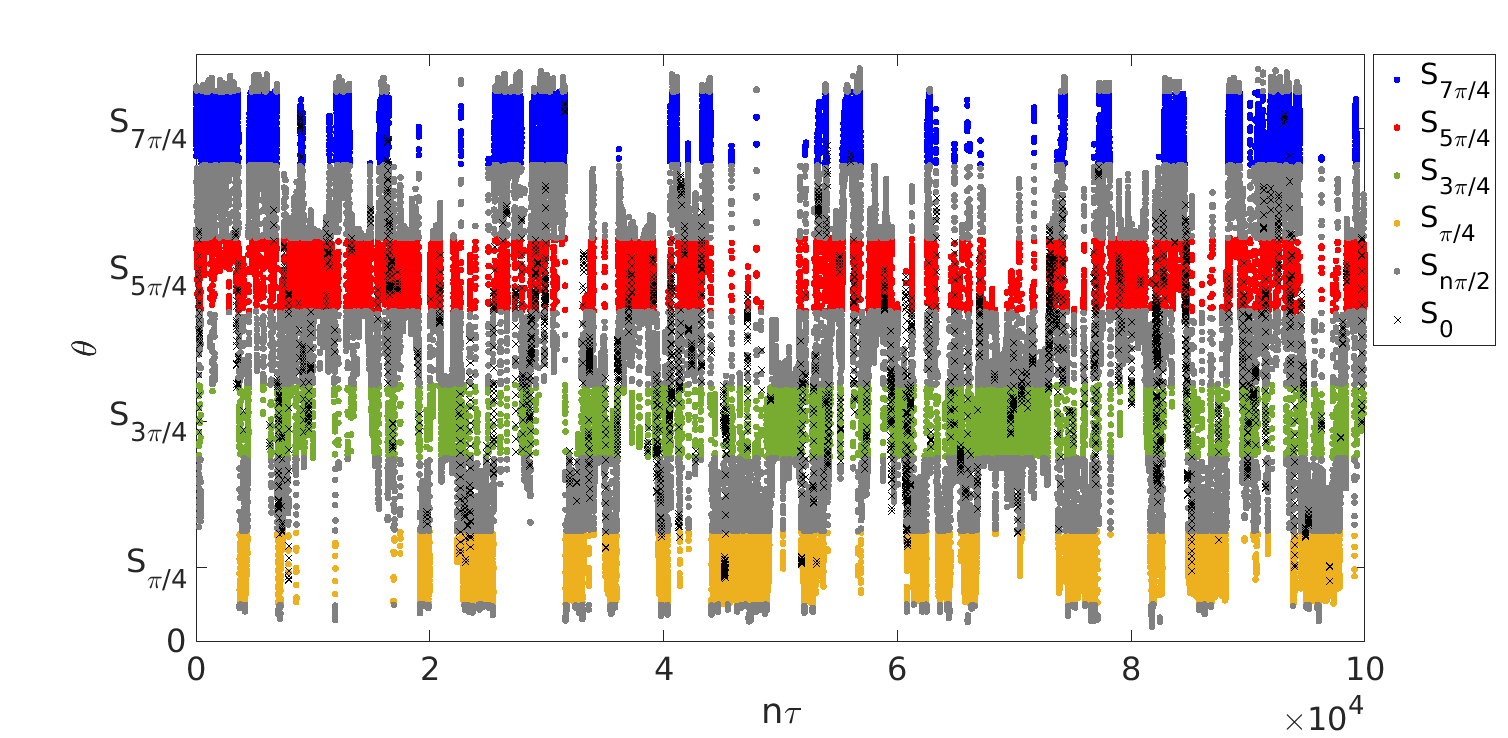}
\caption{Trajectories of the orientation angle $\theta$ of the large-scale circulation for $Pr = 0.1$ (top panel) and $Pr = 0.7$ (bottom panel). Time $t$ is measured in units of the free-fall time $t_f$. States S$_{n\pi/2}$ are given in gray, Null states S$_0$ in black. In both cases, an sliding time-windowed average with a width of 20 $t_f$ is applied. } 
\label{LSC}
\end{figure*}

Figures~\ref{LSC}(a) and (b) show the temporal evolution of the LSC orientation angle ($\theta$) for the cases of $Pr = 0.1$ and $Pr = 0.7$, respectively. The plots underline that the runtime is long enough to capture all 6 states equally. The first four macrostates correspond to  long-lived LSC orientations parallel to the diagonals, with an orientation angle of $\theta = \pi/4, 3\pi/4, 5\pi/4$, and $7\pi/4$ and values of  $\delta \geq \mu - 2\sigma$. All LSC orientations which have $\theta$ values $\pm \pi/8$ to the diagonals are summarized in one macrostate. Therefore, a particular macrostate can have many specific realizations in the system trajectory. The fifth macrostate is the  short-lived LSC state parallel to the edges with $\theta = n \pi/2 \pm \pi/8$ for $n=0, 1 , 2, 3$ and  $\delta \geq \mu - 2\sigma$. We do not distinguish between different $n$ here and summarize all LSC states that are aligned with the side faces into $S_{n\pi/2}$. The final macrostate S$_{0}$ is indicated by the black dots in both panels. Its occurrence is rare as seen in the figure. 

For both Prandtl numbers in figure \ref{LSC}, the stable  long-lived LSC states appear to occur approximately equal number of times. The overall pattern differs however slightly. The case $Pr=0.7$ can be characterized by less frequent switches between the stable states. For the lowest Prandtl number in our series at $Pr=0.1$, the large-scale flow indicates a stronger decoherence for the present Rayleigh number. One possible reason could be the difference of the mean viscous and thermal boundary layer thicknesses, which drive the thermal plume formation and thus LSC jointly. The derivation of clear trends in the Prandtl number dependence of the LSC in this specific geometry would require a generation of additional long-term trajectories at further higher Rayleigh numbers. 

The  short-lived LSC states parallel to the edges, $S_{n\pi/2}$ are seen to fill up the trajectory in both cases as the transition to any  LL-LSC state occurs always via these  SL-LSC states. It is also noted that the data in figure~\ref{LSC} are averaged over 20 time frames to obtain a lucid representation. The Null state appears when there is a transition from one state to another and the flow re-orients itself. {We did not observe a cessation in our records which would imply for example that S$_{\pi/4}$ switches directly to S$_{5\pi/4}$.} In the case of $Pr = 10$, the flow reaches a long-lived state and remains locked there for the rest of the simulation. We followed this state for $5\times 10^4$ free-fall times without a switch. Therefore, this trajectory at the highest of the three Prandtl numbers in our DNS series is not shown here.   

In the following, we will discuss an alternative analysis approach to the large-scale flow in confined turbulent convection, namely by the application of Markov State Models. In such an approach, the main objective is to calculate the macrostate transition probability matrix. Using this information, we can predict how the large-scale statistics of the system evolves in the future, \emph{without} requiring to observe a trajectory that reaches statistical equilibrium.

\section{Markov State Model analysis of the large-scale flow}   
\begin{figure*}
\centering\includegraphics[width=0.9\textwidth]{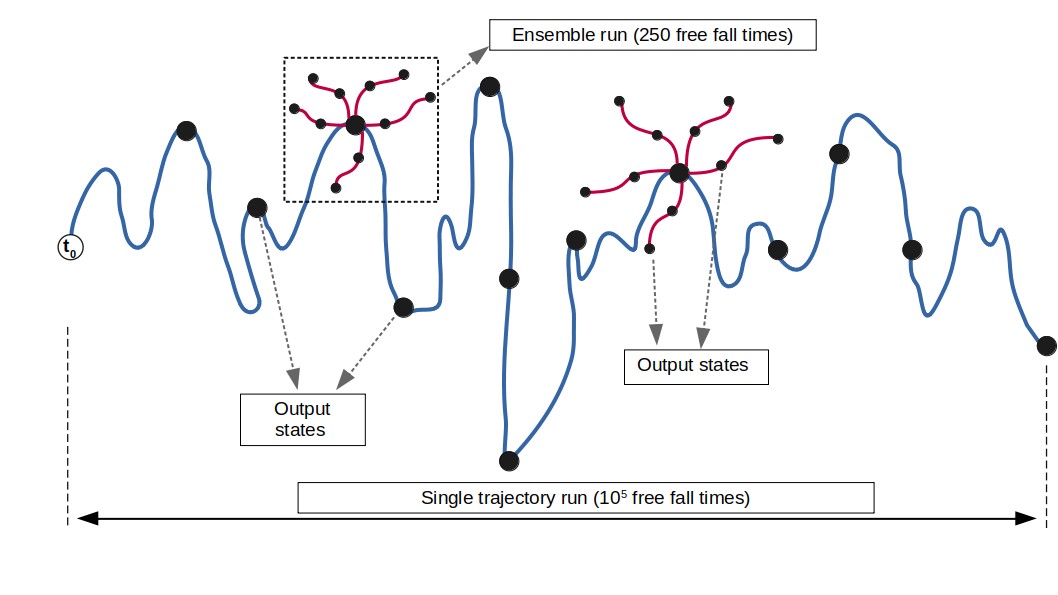}
\caption{Schematic representation of the ensemble averaging process to obtain the transition probability matrix. The blue line denotes long-term single trajectory run, such as the ones shown in figures~\ref{LSC}(a) and (b). The output
states (filled circle, $\bullet$) are printed at every free fall time. The curves in the dashed box represent shorter ensemble runs which start from selected output point of the long-term trajectory. The short-term runs are initially slightly perturbed and evolve in time subsequently. Since the turbulent flow is highly chaotic, differently perturbed short-term runs evolve differently.}
\label{msm}
\end{figure*}
\subsection{Ensemble run setup}
We proceed with the determination of the probability of a transition from one LSC state to another. This analysis is conducted for the case of $Pr = 0.7$ only. The ensemble averaging approach is more prudent than the single trajectory approach, as it mitigates inaccuracy that might arise due to observation of rare events for a single trajectory run. For direct numerical simulations at higher Rayleigh numbers, it might be the only way to sample the phase space sufficiently well, namely to advance with a coarse grid run for a long time interval and to start short-term fully-resolved simulations from the coarse grid outputs. In the present case, the long-term single trajectory run was already sufficiently well resolved such that an upscaling of the short-term runs is not necessary.  
 
As mentioned in the preceding section, we classified the system in six basic states -- four  long-lived LSC configurations $S_{(2n+1)\pi/4}$, the  short-lived LSC configuration $S_{n\pi/2}$, and the Null state~S$_0$. Consequently, a transition probability matrix will be of dimension $6 \times 6$, with the matrix elements $A_{ij}$ representing the probability of transition from state $i$ to state~$j$. {If we aggregate the 4 long-lived macrostates into a single LSC state a $3 \times 3$ transition probability matrix results.} The accuracy of the computed matrix will depend upon two pivotal parameters that we choose: (i) The number of short term simulations for an ensemble averaging, (ii) the duration of a single ensemble run for determining the transition probability. 
{An important, system-intrinsic, factor which severely impairs the accuracy of estimation is the small amount of transitions into rarely visited states (like our Null state). This problem is known in the context of \emph{rare event sampling} and the ensemble setup here is a simple first approach to mitigate it.}
Also, for our method, we have to stick to a finite and attainable amount of runs for ensemble averaging. Recall that each run is a fully resolved DNS. 

\begin{figure*}
\centering
\includegraphics[width=\textwidth]{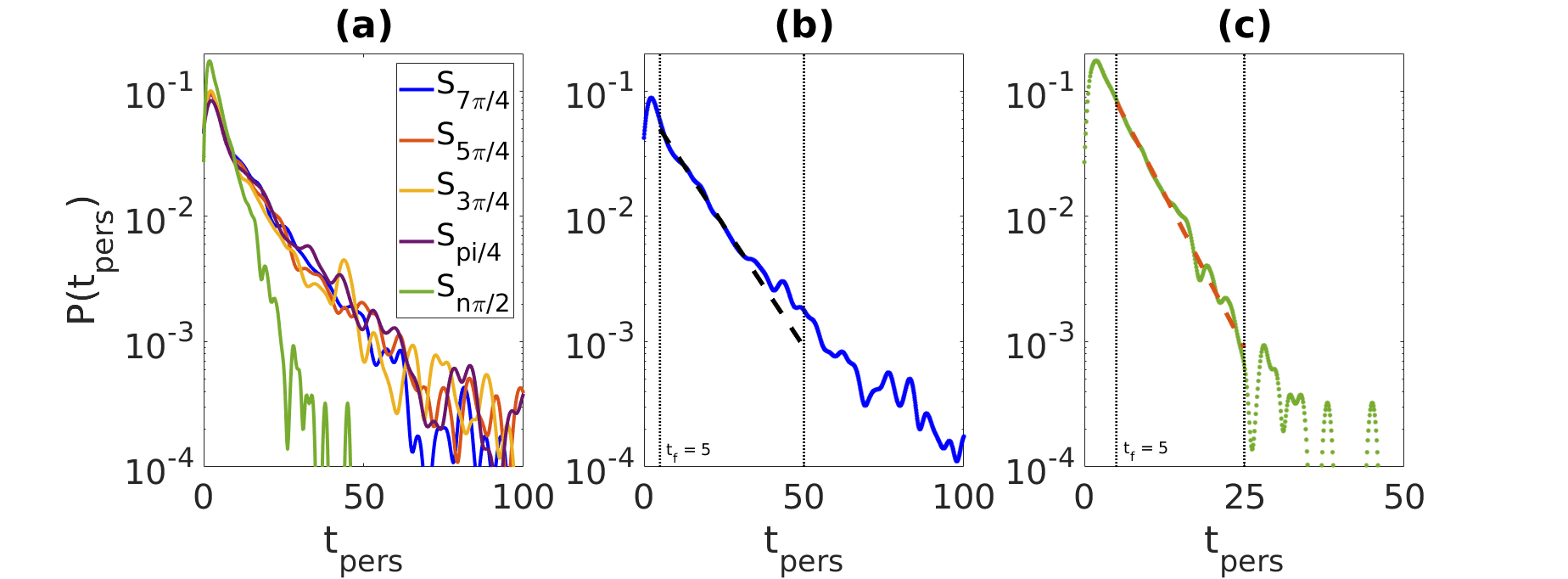}
\caption{{Probability density function (PDF) of the persistence time $t_{\rm pers}$. (a) Separate evaluation for 4 long-lived (LL-LSC) and the short-lived (SL-LSC) macrostates. (b) All 4 long-lived states are combined into one PDF. The exponential fit and the time interval of the fit are indicated by dashed and dotted lines, respectively. (c) Replot of the short-lived macrostate together with the exponential fit and fit interval (as in panel (b)). All data are taken from the long-term trajectory for $Pr = 0.7$.}}
\label{lag_time}
\end{figure*}

The initial conditions for the short-term ensemble simulations were chosen randomly from the output of the single long-term trajectory (as depicted in figure~\ref{msm}). We chose 120 turbulent flow configurations for each specific LSC configuration (macrostate). This results in 720 initial configurations that were picked along the single long-term trajectory in total. Each of the 720 initial conditions is then perturbed five times distinctively. Therefore, in effect, we have 600 of such short-term DNS starting from each LSC configuration or macrostate. The ensemble analysis is consequently based on a total of 3600 short-term trajectories. Unlike the case of single trajectory approach, the output is now coarser with respect to time. The turbulence fields are written out after a finite lag time and not at every free-fall time in order to keep the amount of data limited. 

\subsection{Distribution of persistence time in a macrostate}
The appropriate choice of the output time along the trajectories has to be determined first. If the lagtime is too small then one would not observe transitions to another LSC state. For long lagtimes, one would miss information on intermediate transitions. Also, the longer the lagtime, the larger the ensemble to resolve the transition statistics, as it spreads over more macrostates. To evaluate the lagtime, we determine the time for which the system remains in a state before crossing over to the next state and label this time as the persistence time $t_{\rm pers}$. The persistence times $t_{\rm pers}$ for all possible stable long- and short-lived LSC states were computed separately, using the data from the original long-term single trajectory run. The rarely occurring Null states were excluded from this analysis. 

The probability density functions (PDFs) of these persistence times are shown in figure~\ref{lag_time}(a). The PDFs for all 5 macrostates peak around $5t_f$, indicating that for most initial conditions a first transition can occur after such a short time interval. These PDFs are obtained using the raw data without any averaging and hence this short persistence time is connected to the turbulent fluctuations of the convection flow. Note also that all 4 PDFs for the LL-LSCs have extended tails up to 100 free fall times. The latter time would correspond to the persistence time of a macrostate, if we average out small scale fluctuations, as observed from figure~\ref{LSC}. The PDF of the SL-LSC decays more rapidly as visible in figure~\ref{lag_time}(a). 

{An estimate of the mean lifetime of the macrostates is obtained by fitting an exponential function to the tails of the PDFs. Figure~\ref{lag_time}(b) therefore aggregates the 4 LL-LSC states into one state with a corresponding PDF.  The distribution of persistence times follows an exponential distribution (within the time interval $5\le t_f \le 50$, as indicated by the dashed line). A similar behavior was observed in previous studies of other types of convection flows~\cite{mishra_2011, verma_2015, mannatil_2017}. The reciprocal of the exponential fit coefficient gives a mean lifetime of the LL-LSC states of about~12$t_f$. This is the time scale before a spontaneous transition to either the SL-LSC or the Null state. A similar exponential fit is reported in figure~\ref{lag_time}(c) to the persistence time distribution of the SL-LSC. The time interval for the fit is now $5\le t_f \le 25$ as indicated again by the dashed line. The mean lifetime of the SL-LSC state follows to about 4.2$t_f$. The LL-LSC states persist almost three times as long as the SL-LSC state, marking them as the most probably observed macrostates along the long term trajectory. This information explains the rationale behind providing utmost weightage to the LL-LSC state in generating the Markov State Model.}

\begin{figure*}
\centering\includegraphics[width=0.8\textwidth]{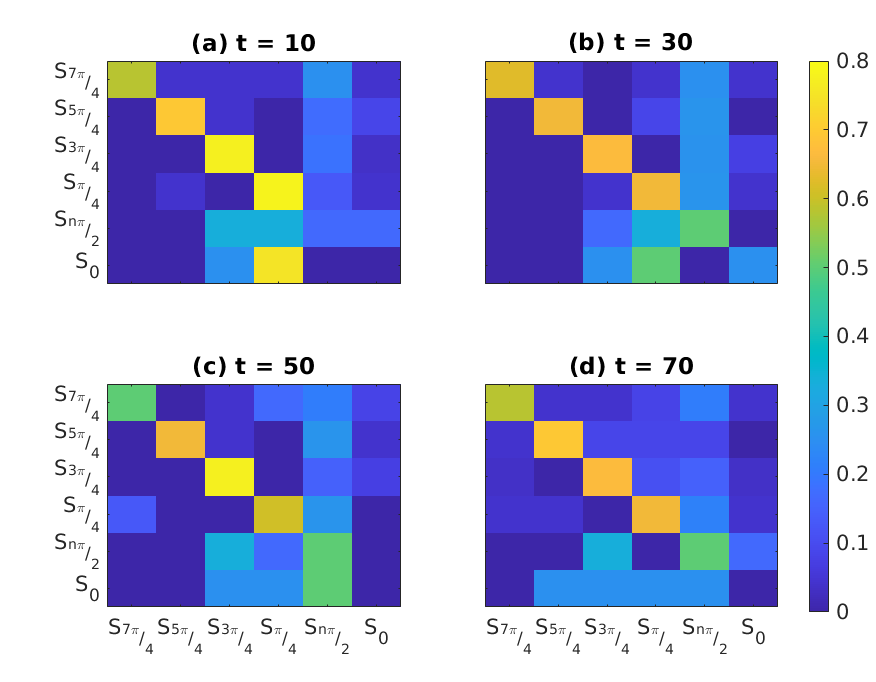}
\caption{Time evolution of the transition probability matrix. The color coded transition probability matrix are plotted for four different values of time instants (a) $t = 10$, (b) $t = 30$, (c) $t = 50$, and (d) $t = 70$ free fall time units.} 
\label{trans_prob_matrix}
\end{figure*}
\begin{figure*}
\centering\includegraphics[width=0.98 \textwidth]{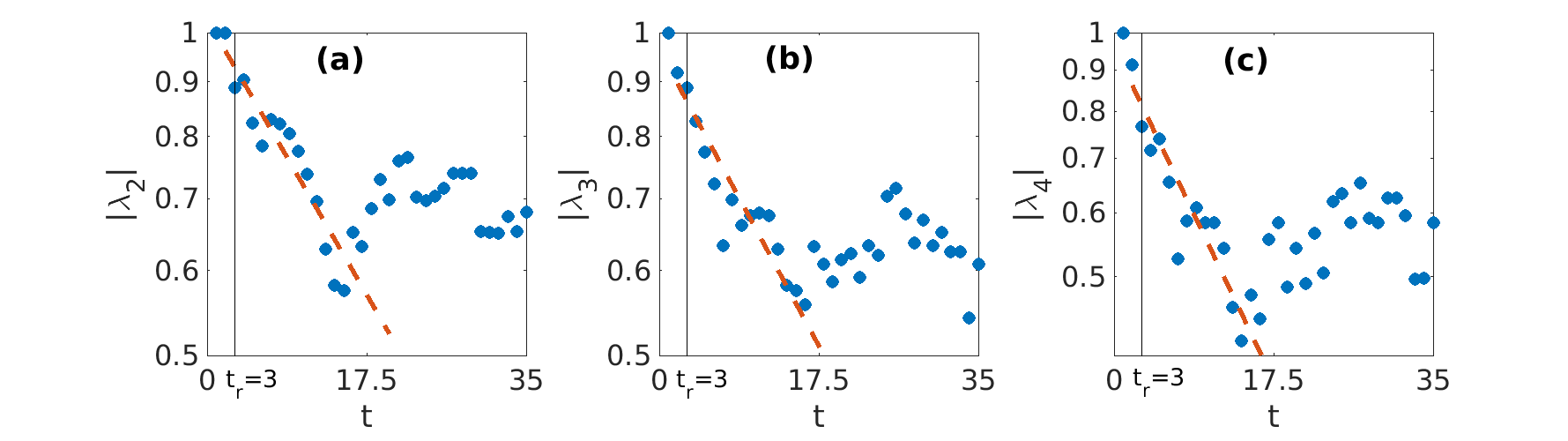}
\caption{Plot of absolute values of second, third, and fourth-largest eigenvalues of the transition probability matrix as a function of time (blue dots). The eigenvalues are computed from the matrices $\mathbf{A}_t$ for each time frame and the red dashed curve represents the fitted exponential curve. The data for the fit are the eigenvalues for~{$20 > t > t_r = 3$}.} 
\label{eigenvalues}
\end{figure*}

\subsection{Transition probabilities for 3-state and 6-state models}
{We proceed in our analysis by taking a lagtime of 10$t_f$ which is twice the most probable persistence time $t_{\rm pers}$. Thus, more than $50\%$ of the cases would transition to the subsequent macrostate assuming that the states get uncorrelated after a time of about 10$t_f$. We used again the method described in section~\ref{LTS} to identify the state of the system at the corresponding time and to determine the components
\[
A_{ij} = \mathrm{Prob}\left[ \text{system is in macrostate $j$ at time $t_0+t$} \mid \text{system is in macrostate $i$ at time $t_0$} \right]
\]
of the transition matrix~$\mathbf{A}_t$ corresponding to the lagtime~$t$. For a time-homogeneous process (that we assume to have here) the initial time $t_0$ is arbitrary. Averages are now taken however over the ensemble of 3600 trajectories.
}

{Figure~\ref{trans_prob_matrix} represents the color-coded transition probability matrix along the short-term trajectories of 250$t_f$ for the 6 macrostates. Each box in the figure signifies the value of a component $A_{ij}$ of the matrix. The four panels of the figure show that the probabilities change slightly with changing lagtime.}
For the diagonal flow configurations, self-transitions appear to be the most likely transitions. The transition probability is lowest to the $S_0$ state. As can be also seen, the data is still accompanied by considerable noise. 
An ensemble of 3600 trajectories still seems be too small.

{Table \ref{prob_matrix_3} repeats the analysis for three macrostates and shows the results at three time instants in (a--c). This implies that now all four LL-LSC states are summarized into one long-lived state, which is termed S$_L$ in the following. This switch to one macrostate can be done for symmetry reasons in the flow. Thus we probe transitions between S$_L$, S$_{n\pi/2}$, and S$_0$.  The data are listed now in the form of small tables for better visibility, each for the lagtime that is indicated at the top. 
An estimate of the statistical weights of different states 
can be calculated using the normalized eigenvector corresponding to the first eigenvalue. The weightages of different states, corresponding to the three-macrostate model are enlisted in table~\ref{prob_matrix_3}(d). We observed the system to rest in an LL-LSC state for about 75\% of the time of the ensemble runs, a fact also reflected by the PDFs of persistence times.}
\begin{center}
\begin{table}[h]
    \begin{subtable}[h]{0.45\textwidth}
    \centering
        \caption{t = 5$t_f$}
        \begin{tabular}{c|ccc}
         & $S_{L}$ & $S_{n\pi/2}$ & $S_0$ \\
        \hline 
        $S_{L}$ &0.702 & 0.252 & 0.040 \\
        $S_{n\pi/2}$ & 0.519 & 0.481 & 0.0 \\
        $S_0$ & 0.761 & 0 & 0.239
       \end{tabular} 
    \end{subtable}
    \hfill
    \begin{subtable}[h]{0.45\textwidth}
    \centering
        \caption{t = 20$t_f$}
        \begin{tabular}{c|ccc}
         & $S_{L}$ & $S_{n\pi/2}$ & $S_0$ \\
        \hline 
       $S_{L}$ &0.754 & 0.205 & 0.041 \\
       $S_{n\pi/2}$ &1.0 & 0.0 & 0.0 \\
       $S_0$ &0.240 & 0.510 & 0.250
       \end{tabular}
     \end{subtable} 
     \\
     \begin{subtable}[h]{0.45\textwidth}
     \centering
        \caption{t = 40$t_f$}
        \begin{tabular}{c|ccc}
          & $S_{L}$ & $S_{n\pi/2}$ & $S_0$ \\
        \hline 
        $S_{L}$ &0.756 & 0.222 & 0.022 \\
       $S_{n\pi/2}$ &0.334 & 0.333 & 0.333 \\
       $S_0$ &1.0 & 0.0 & 0.0
       \end{tabular}
     \end{subtable}
     \hfill
     \begin{subtable}[h]{0.45\textwidth}
     \centering
        \caption{weightage of states}
        \begin{tabular}{c|ccc}
          & $5t_f$ & $20 t_f$ & $40 t_f$ \\
        \hline 
       $S_{L}$ &0.763 & 0.776 & 0.682 \\
       $S_{n\pi/2}$ &0.169 &0.181 & 0.227  \\
       $S_0$ &0.068 & 0.043 & 0.091
       \end{tabular}
     \end{subtable}
     
     \caption{  Evolution of the three-macrostate model: The transition probability matrices of the three-macrostate model at (a) t = 5$t_f$, (b) t = 20$t_f$, and (c) t = 40$t_f$. The states S$_L$, S$_{n\pi/2}$, and S$_0$ represents the long-lived LSC, short-lived LSC, and the Null state respectively. The weightage of different states according to their probability of occurrence at different time instants is given in table (d).}
     \label{prob_matrix_3}
\end{table}
\end{center}

\subsection{Markov property by eigenvalue spectra}
\label{ssec:Markov_eig}

Based on the ensemble run data, we now proceed to test the Markovianity of the macrostate evolution. This is done by means of the eigenvalue spectrum of the transition probability matrices $\mathbf{A}_t$ at different lagtimes $t$ as discussed in the last section. To begin with, we calculated the ensemble averaged transition probability matrix $\mathbf{A}_t$ at different lagtimes $t \ge t_r$ where $t_r=3 t_f$ is chosen as a reference time for Markov analysis. The eigenvalues~$\lambda_i(t)$, $i=1,\ldots,6$ are always ordered according to descending magnitude. The largest eigenvalue is $\lambda_1=1$ at all times, since the matrices are stochastic. The subsequent eigenvalues may remain unity for times $t \lesssim t_r$; afterwards {the trajectories} start to leave the initial macrostates and {the eigenvalues} decrease thereafter.

If $\mathbf{A}_t$ defines the transition matrix up to lagtime $t$ then for a truly Markovian process the so-called {\em Chapman--Kolmogorov identity} should be satisfied. It is given by 
\begin{equation}
\mathbf{A}_{t+s} = \mathbf{A}_t \mathbf{A}_s\quad\mbox{for}\quad s,t>0\,.
\end{equation}
This means for the eigenvalues, that they must satisfy~$\lambda_i(t) = \exp(-\omega_i t)$ for some \emph{rate}~$\omega_i \in \mathbb{C}$. Clearly, due to the coarse graining, as we discussed above, we can expect the Chapman--Kolmogorov identity to hold best for lagtimes $s,t \ge t_r$. Less dominant, i.e., faster decaying, eigenvalues correspond to faster dynamical processes, hence they are harder to estimate by the sampling procedure. We thus discard the two smallest eigenvalues $\lambda_5$ and $\lambda_6$ and test $|\lambda_i(t)| \approx \exp(-\omega_i t)$ for $t > t_r$ for $i=2,3,4$, and appropriate~$\omega_i >0$.

For $t>t_r=5$ we output each 10 free-fall times. These data are used to test Markovianity.  The trend of variation of eigenvalues $\lambda_i(t)$ of $\mathbf{A}_t$ as a function of time yielded in a first analysis strong fluctuations, most likely due to the insufficient sample size. Hence, to obtain superior statistical averaging, we further sub-divided the short-term ensemble DNS into three different sections of a temporal length of 70 free-fall time each, separated by a time window of $10 t_f$ in order to decorrelate the subsequent sections in the analysis. The three data sets, which are obtained in this way, were treated as independent trajectories. This includes the whole process of computing the matrices $\mathbf{A}_t$ and their eigenvalues $\lambda_i(t)$. 

Figure~\ref{eigenvalues} shows the magnitude of the second, third, and fourth-largest eigenvalues of $\mathbf{A}_t$ as a function of lagtime. Note that for these fits eigenvalues were computed at every free-fall time.  {The eigenvalues show a qualitatively superior fit to an exponential decay for $t \lesssim 20 t_{f}$. Beyond this time, the eigenvalues fluctuate about a constant mean. This indicates that the system exhibits Markovianity for times~$t \lesssim 20 t_{f}$, but the ensemble statistics is still insufficient to conclude Markovianity for larger time intervals. This statement was tested by re-running the analysis with 1800 trajectories only. An increase in the size of the ensemble will thus lead to an extension of the time interval for which the eigenvalues show an exponential decay.} 

\begin{figure*}
\centering\includegraphics[width=0.6\textwidth]{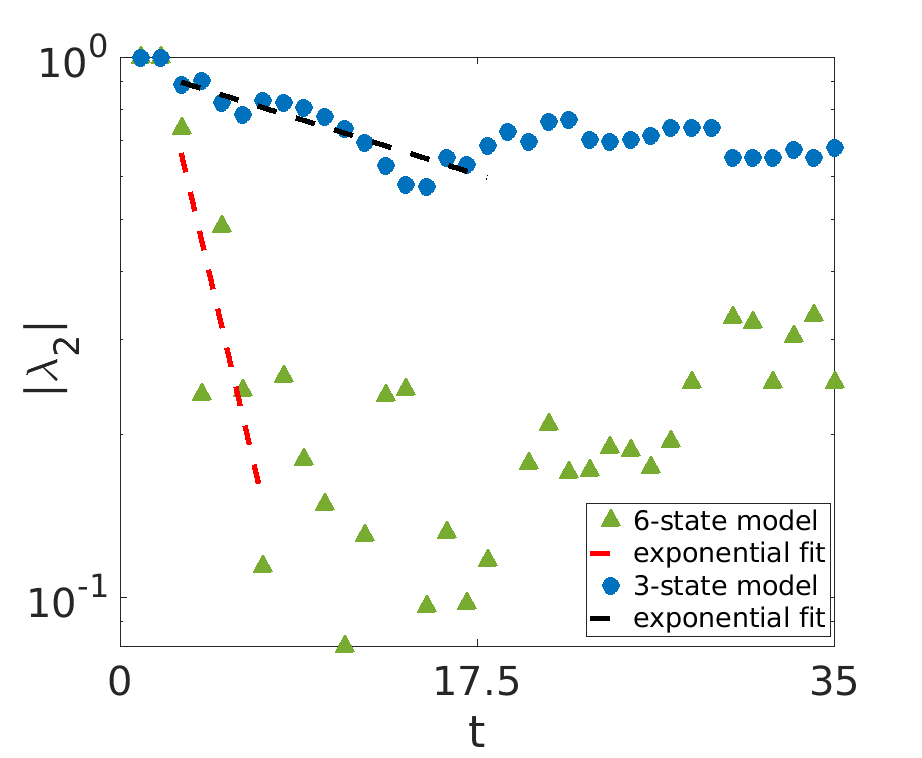}
\caption{Comparison of the eigenvalue spectrum of the three-state and six-state model. The second largest eigenvalue of the transition probability matrix $A_{ij}$ as a function of time is computed for both models. The dashed lines are the fits to the exponential decay.} 
\label{three-state-model}
\end{figure*}


{Figure~\ref{three-state-model} shows a comparison of the magnitude of the second eigenvalue versus time for the three-state and the six-state models. The plot shows that the results are now even stronger affected by statistical uncertainties. The second eigenvalue is closer to an exponential decay in case of the six macrostates. This validates the advantage of the six-state model. A simple qualitative explanation for the observed behavior in the three-state model could be as follows. Consider for simplicity the long-lived (LL) state S$_L$ and short-lived (SL) state S$_{n\pi/2}$ only. Let $0<a,b\ll 1$ be two small parameters such that the transition probability matrix between these two macrostates is given by 
\begin{displaymath}
    \left(
\begin{array}{ll}
\rm LL\to \rm LL & \rm LL \to \rm SL \\
\rm SL \to \rm LL & \rm SL \to \rm SL \\
\end{array}\right) = 
    \left(
\begin{array}{ll}
1-a & a \\
1-b & b \\
\end{array}\right)\,. 
\end{displaymath}
The second eigenvalue of this matrix is $b-a$, with~$|b-a|\ll 1$. The strong drop in magnitude is what we see for the spectrum of the three-state model.}
{By aggregating the LL-LSC to obtain the three-state model we eliminate exactly the long timescales from the six-state process. This gives us a process with only one dominant state, where the other states are left very quickly by the process in comparison with the six-state one, mimicked by the small parameter~$b$ above. This imbalance of persistences is more difficult to estimate, yielding the large statistical errors in the second eigenvalue of the three-state model.}

\begin{figure*}
\centering\includegraphics[width=\textwidth]{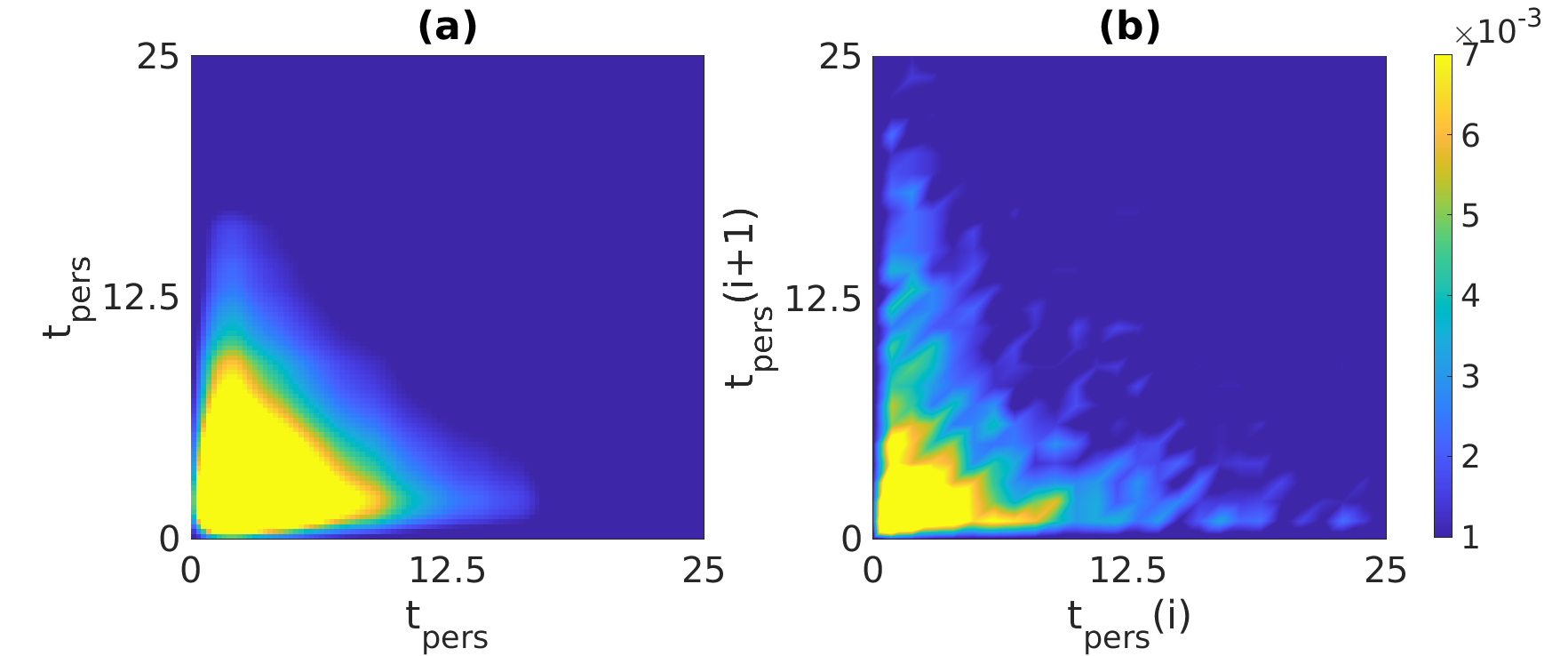}
\caption{{Comparison of joint probability density functions (JPDFs) of persistence times $t_{\rm pers}$. (a) Heat map of the JPDF $p(t_{\rm pers}) \otimes p(t_{\rm pers})$, which is obtained from outer product of the PDFs that are shown in figure \ref{lag_time}(b). (b) Heat map of JPDF $p(t_{\rm pers}(i), t_{\rm pers}(i+1))$ where $t_{\rm pers}(i)$ is the persistence time of the $i$-th member of the macrostate time series. The color bar holds for both panels.}} 
\label{joint}
\end{figure*}

\subsection{Markov property by the persistence time distributions}
\label{ssec:Markov_pers}

{As shown in figure~\ref{lag_time}, the PDFs of the persistence times $t_{\rm pers}$ of the LL-LSC and SL-LSC display an exponential decay. Consequently, we could conclude the Markov property if consecutive jumps would be independent. To test this, we compare the distribution $p(t_{\rm pers}(i)), p(t_{\rm pers}(i+1))$ of consecutive jump times with the (so called ``outer'') product of the jump time distribution with itself,~$p(t_{\rm pers}) \otimes p(t_{\rm pers})$. The latter can be thought of as the joint probability distribution of an independent pair of persistence times. Markovianity is then satisfied when these two distributions are equal, i.e., when two successive jumps between macrostates are independent.  Figure~\ref{joint} shows both distributions and compares them directly. It can be concluded that this property is satisfied fairly well. This short analysis thus provides a further indication of the Markovianity of the large-scale flow dynamics in the present turbulent flow.

We remark that for fully concluding Markovianity, both the destination and the jump times of consecutive jumps need to be mutually independent for \emph{all individual} states. As the short-lived state and in particular the Null state are much quicker exited than the long-lived one, the associated statistics is less revealing. Hence, we restrict our attention here to the somewhat weaker analysis of persistence times of all states in an aggregated manner.
}

\section{Summary and conclusion}
The present numerical study of turbulent Rayleigh--B\'{e}nard convection was focused on the large-scale flow behavior in a closed cubic cell, a possible configuration of laboratory measurements. Contrary to the more frequently applied cylinder configuration the present geometry lacks an azimuthal symmetry, which generally causes a slow drift of the large-scale circulation configuration as a whole, see e.g. ref. \cite{zuerner_2019}. {The constrained flow in the cube can be characterized by several {\em discrete} macrostates. These are either 3 or 6 large-scale flow configurations. In the former case, one summarizes the 4 diagonal LSC configurations to one representative long-lived state for symmetry reasons. Turbulent convection generates an approximate hopping dynamics between these different states in the high-dimensional phase space given that the Prandtl number is sufficiently low at a given Rayleigh number (which was here $Ra=10^6$).} In this respect, the RBC flow behaves similarly to molecular dynamics systems, such as proteins, that undergo rare and rapid conformational changes. This motivated the application of a Markov State Model to the RBC system in this work. 

First, we studied the dynamics along a single long-term trajectory that extends over $10^5$ free-fall time units $t_f$. Along this trajectory each of the macrostates appeared sufficiently often which allowed us to determine their characteristic  lifetime. It is found that this time depends strongly on the Prandtl number. For $Pr\lesssim 1$ our system switches rapidly between different macrostates. For $Pr\gg 1$, the LSC remains locked in one particular configuration for $5\times 10^4 t_f$. This behavior will depend on the Rayleigh number, which was fixed to one value only here. As can be seen in Table \ref{tab1}, the resulting momentum transfer at $Ra=10^6$ and $Pr=10$, which is quantified by the Reynolds number $Re$ remains too small. The flow thus lacks sufficient fluctuations of the velocity field that are able to transfer the flow into a new configuration. For the case of $Pr=0.7$, we analysed the transitions between the different macrostates in the form of an ensemble analysis and determined the corresponding transition probability matrix. It was shown that the elements of this matrix depend strongly on the lagtime that is used to probe a transition from a macrostate $i$ to a macrostate~$j$.

A subsequent analysis of the Markov State Model revealed that the ensemble size of 3600 short-term DNS used to assemble the model might be still too small and statistical errors still affect dynamical properties. However, the transition probability matrix did show Markovian properties for times $t \lesssim 20 t_f$, which we derived from the eigenvalue spectrum in the 6-state model. Further, sampling beyond the one along the 3600 trajectories might be necessary to assess how a faithful Markov model could be constructed.
{An equivalent characterization of a Markov (jump) process is by the means of exponential persistence (or ``holding'') time distributions and independence of consecutive jumps. The quantitative estimation of these properties turned out to be a more viable task for our setting than faithful componentwise estimation of the transition matrix for a fixed lagtime.
We demonstrated that the persistence times of the LSCs follow an exponential distribution and also demonstrated the ``memoryless property'' here fairly well by means of the PDF of the persistence time.
}


{From the perspective of the analysis}, one could question the choice of macrostates, and methods could be considered that estimate surrogate macrostate variables directly from data. These are either tuned to give optimal representation of the state-space geometry~\cite{GHB15,giannakis_2018,KoWe18}, or to reproduce optimally certain statistical properties of the coarse-grained dynamics~\cite{BitEtAl17,PBLRS20}.  In the present work, the simpler 3-state model, which we included in a our analysis, did not lead to an improvement in the results.

Our study is thought as a first proof of concept. {From the perspective of the system under consideration}, several routes are possible to extend our present research. It would be desirable to increase the Rayleigh number to values of $10^7$ or $10^8$ which would be line with a modified hopping dynamics. We expect less coherent large-scale flows due to the enhanced level of velocity fluctuations and thus shorter mean lifetimes for the macrostates and for all Prandtl numbers. A second route would follow an analysis in larger-aspect-ratio configurations, i.e., for convection in flat domains. The switching between different large-scale patterns that might contribute differently to the heat transfer would be an interesting task for future work. {The latter application could then pave the way to new parametrisation models of mesoscale convection.}   

\section*{Authors' contributions}
P.K. and J.S. designed the research. P.M. conducted the simulations and analyzed the data. All three authors discussed the results and drafted the manuscript. All three authors gave final approval for publication.

\section*{Competing Interests}
We declare we have no competing interests.

\section*{Funding}
Priyanka Maity is supported by the grants SCHU 1410/29-1 and SCHU 1410/30-1 of the Deutsche Forschungsgemeinschaft (DFG).
P\'eter Koltai has been partially supported by the DFG through grant CRC 1114 ``Scaling Cascades in Complex Systems'', Project Number 235221301, Project A01 ``Coupling a multiscale stochastic precipitation model to large scale atmospheric flow dynamics’' and and under Germany's Excellence Strategy -- The Berlin Mathematics Research Center MATH+ (EXC-2046/1 project ID: 390685689).

\section*{Acknowledgements}
The comprehensive ensemble simulations were conducted at the MaPaCC4 compute cluster of the University Computing Centre at Technische Universit\"at Ilmenau (Germany).

\section*{Disclaimer}
Not applicable.



\begin{thebibliography}{100}

\bibitem{spiegel_1971}
Spiegel EA. 1971 Convection in stars i. Basic Boussinesq convection.
\textit{Annu. Rev. Astron. Astrophys.} \textbf{9}, 323--352.

\bibitem{marshall_schott_1999}
Marshall J, Schott F. 1999 Open-ocean convection: Observations, theory, and models.
\textit{Rev. Geophys.} \textbf{37}, 1--64.

\bibitem{markson_1975}
Markson R. 1975 Atmospheric electrical detection of organized convection.
\textit{Science} \textbf{188}, 1171--1177.

\bibitem{hathaway_2012}
Hathaway DH. 2012 Supergranules as probes of solar convection zone dynamics.
\textit{Astrophys. J. Lett.} \textbf{749} L13.

\bibitem{schumacher_2020}
Schumacher J, Sreenivasan KR. 2020 Colloquium: Unusual dynamics of convection in the Sun. 
\textit{Rev. Mod. Phys.} \textbf{92} 041001.

\bibitem{chandrashekhar_1961}  Chandrashekhar S. 1961  \textit{Hydrodynamic and Hydromagnetic instability}. Dover Publication, UK.

\bibitem{ahlers_2009}
Ahlers G, Grossmann S, Lohse D. 2009 Heat transfer and large scale dynamics in turbulent Rayleigh-B\'enard convection. \textit{Rev. Mod. Phys.} \textbf{81}, 503--537.

\bibitem{chilla2012}
Chill\`{a} F, Schumacher J. 2012 New perspectives in turbulent Rayleigh-B\'{e}nard convection. Eur. Phys. J. E \textbf{35}, 58.

\bibitem{pandey_2018}
Pandey A, Scheel JD, Schumacher J. 2018 Turbulent superstructures in Rayleigh-B\'{e}nard convection. Nat. Commun. \textbf{9}, 2118. 

\bibitem{niemela_2001}
Niemela JJ, Skrbek L, Sreenivasan KR, Donelly RJ. 2001 The wind in confined thermal convection.
\textit{J. Fluid Mech.} \textbf{449}, 169--178.

\bibitem{sreenivasan_2002}
Sreenivasan KR, Bershadskii A, Niemela JJ. 2002 Mean wind and its reversal in thermal convection.
\textit{Phys. Rev. E} \textbf{65}, 056306.

\bibitem{parodi_2004}
Parodi A, von Hardenberg J, Passoni G, Provenzale A, Spiegel EA. 2004 Clustering of plumes in turbulent convection. \textit{Phys. Rev. Lett.} \textbf{92}, 194503.

\bibitem{puthenveettil_2005}
Puthenveettil BA, Arakeri JH. 2005 Plume structure in high-Rayleigh-number convection.
\textit{J. Fluid Mech.} \textbf{542} 217--249.

\bibitem{brown_2006}
Brown E, Ahlers G. 2006 Rotations and cessations of the large-scale circulations in
turbulent Rayleigh-B\'enard convection. \textit{J. Fluid Mech.} \textbf{568}, 351--386.
 
\bibitem{xi_2007}
Xi H-D, Xia K-Q. 2007 Cessations and reversals of the large-scale circulations in turbulent
thermal convection. \textit{Phys. Rev. E} \textbf{75} 066307.

\bibitem{Verma_book_2018}  Verma M. K. 2018.  \textit{Physics of Buoyant Flows: From Instabilities to Turbulence}. World Scientific,
Singapore.

\bibitem{mishra_2011}
Mishra P. K., De A. K., Verma M. K., and Eswaran V., 2011 Dynamics of reorientations and reversals
of large-scale flow in Rayleigh–B\'enard convection. \textit{J. Fluid Mech.} \textbf{668} 480.

\bibitem{verma_2015}
Verma M. K., Ambhire S. C., and Pandey A., 2015 Flow reversals in turbulent convection with free-slip
walls. \textit{Phys Fluids.} \textbf{27} 047102.

\bibitem{mannatil_2017}
Mannattil M., Pandey A., Verma M. K., and Chakraborty S.,2017 On the applicability of low-dimensional
models for convective flow reversals at extreme Prandtl numbers. \textit{Eur. Phys. J. B.} \textbf{90} 259.

\bibitem{kumar_2018}
Kumar A., and Verma M. K., 2018 Applicability of Taylor’s hypothesis in thermally driven turbulence. \textit{Royal Soc. Open Sci.,} 
\textbf{5} 172152.

\bibitem{bao_2015}
Bao Y, Chen J, Liu B-F, She Z-S, Zhang J, Zhou Q 2015 Enhanced heat transport in 
partitioned thermal convection. \textit{J. Fluid Mech.} \textit{784} R5.

\bibitem{wagner_2013}
Wagner S, Shishkina O. 2013 Aspect-ratio dependency of Rayleigh-B\'enard convection in a box shaped containers. \textit{Phys. Fluids} \textbf{25} 085110.

\bibitem{chong_2015}
Chong KL, Huang S-D, Kaczorowski M, Ni R, Xia K-Q. 2015 Condensation of coherent structures in
turbulent flows. \textit{Phys. Rev. Lett.} \textbf{115} 264503.

\bibitem{daya_2001}
Daya ZA, Ecke RE. 2001 Does turbulent convection feel the shape of the container?
\textit{Phys. Rev. Lett.} \textbf{87}, 184501.

\bibitem{hartmann_2021}
Hartmann R, Chong KL, Stevens RJAM, Verzicco R, Lohse D. 2021 Heat transport enhancement in confined Rayleigh-B\'{e}nard convection feels the shape of the container.
\textit{Europhys. Lett.} \textbf{135}, 24004.

\bibitem{foroozani_2014}
Foroozani N, Niemela JJ, Armenio V, Sreenivasan KR. 2014 Influence of container
  shape on scaling of turbulent fluctuations in convection.
\textit{Phys. Rev. E} \textbf{90}, 063003.

\bibitem{bai_2016}
Bai K, Ji D, Brown E. 2016 Ability of a low-dimensional model to predict geometry-dependent dynamics of large-scale coherent structures in turbulence. \textit{Phys. Rev. E} \textbf{93}, 023117.

\bibitem{foroozani_2017}
Foroozani N, Niemela JJ, Armenio V, Sreenivasan KR. 2017 Reorientations of the
  large-scale flow in turbulent convection in a cube.
\textit{Phys. Rev. E} \textbf{95}, 033107.

\bibitem{giannakis_2018}
Giannakis D, Kolchinskaya A, Krasnov D, Schumacher J. 2018 Koopman analysis of
  the long-term evolution in a turbulent convection cell.
\textit{J. Fluid Mech.} \textbf{847}, 735–767.

\bibitem{vasilev_2019}
Vasilev A, Frick P, Kumar A, Stepanov R, Sukhanovskii, Verma MK. 2019 Transient flows and reorientations of large-scale convection in a cubic cell.
\textit{Int. Commun. Heat Mass Transfer} \textbf{108}, 104319. 

\bibitem{teimurazov_2021}
Teimurazov A, Reiter P, Shishkina O, Frick, P. 2021 Heat transport in a cell heated at the bottom and the side.
\textit{Europhys. Lett.} \textbf{134}, 34001.

\bibitem{schikarski_2019}
Schikarski T, Trzenschiok H, Peukert W, Avila M. 2019 Inflow boundary conditions determine T-mixer efficiency.
\textit{React. Chem. Eng.} \textbf{4}, 559--568.

\bibitem{mezic_2013}
Mezic I. 2013 Analysis of fluid flows via spectral properties of the Koopman operator.
\textit{Annu. Rev. Fluid Mech.} \textbf{45}, 357--378.

\bibitem{williams_2015}
Williams MO, Kevrekidis IG, Rowley CW. 2015 A data-driven approximation of the Koopman operator: Extending dynamic mode decomposition. 
\textit{J. Nonlinear Sci.} \textbf{25}, 1307--1346.

\bibitem{sauer_1991}
Sauer T, Yorke JA. 1991 Shadowing trajectories of dynamical systems, In: Meyer K.R., Schmidt D.S.
(eds) {\em Computer Aided Proofs in Analysis}. The IMA Volumes in Mathematics and Its Applications, vol
28. Springer, New York, NY, 229-234.

\bibitem{pande_2010}
Pande VS, Beauchamp K, Bowman GR. 2010 Everything you wanted to know about Markov State Models but were afraid to ask.
\textit{Methods} \textbf{52}, 99--105.
 
\bibitem{husic_2018}
Husic BE, Pande VS. 2018 Markov State Models: From an art to a science.
\textit{J. Am. Chem. Soc.} \textbf{140}, 2386--2396.
 
\bibitem{prinz_2011}
Prinz JH, Wu H, Sarich M, Keller B, Senne M, Held M, Chodera JD, Sch\"utte C, No\'e F. 2011 Markov models of molecular kinetics: Generation and validation.
\textit{J. Chem. Phys.} \textbf{134}, 174105.

\bibitem{SchSa13}
Sch{\"u}tte C, Sarich M. 2013 {\em Metastability and Markov State Models in Molecular Dynamics.}
Courant Lecture Notes in Mathematics.

\bibitem{BoPaNo14}
Bowman GR, Pande VS, No\'e F. 2014 
{\em An Introduction to Markov State Models and Their Application to Long Timescale Molecular Simulation} Vol. 797, {\em Advances in Experimental Medicine and Biology}, Springer, Berlin.

\bibitem{SchEtAl15}
Scherer MK, Trendelkamp-Schroer B, Paul F, P\'erez-Hern\'andez G, Hoffmann M, Plattner N, Wehmeyer C, Prinz JH, No\'e F. 2015 PyEMMA 2: A software package for estimation, validation, and analysis of Markov models. \textit{J. Chem. Theory Comput.} \textbf{11}, 5525--5542.

\bibitem{MSMBuilder17}
Harrigan MP, Sultan MM, Hern\'andez CX, Husic BE, Eastman P, Schwantes CR, Beauchamp KA, McGibbon RT, Pande VS. 2017 MSMBuilder: Statistical Models for Biomolecular Dynamics.
\textit{ Biophys. J.} \textbf{112}, 10--15.

\bibitem{KoWe18}
Koltai P, Weiss S. 2020 Diffusion maps embedding and transition matrix analysis of the large-scale flow structure in turbulent Rayleigh--B\'enard convection. \textit{Nonlinearity} \textbf{33} 
1723--1756.

\bibitem{fischer_1997}
Fischer PF. 1997 An overlapping Schwarz method for spectral element solution of the incompressible Navier–Stokes equations.
\textit{J. Comp. Phys.}  \textbf{133}, 84--101.
  
\bibitem{scheel_2013}
Scheel JD, Emran MS, Schumacher J. 2013 Resolving the fine-scale structure in turbulent Rayleigh-B\'{e}nard convection. \textit{New J. Phys.} \textbf{15}, 113063.  

\bibitem{zuerner_2019}
Z\"urner T, Schindler F, Vogt T, Eckert S, Schumacher J. 2019 Combined measurement of velocity and temperature in liquid metal convection. \textit{J. Fluid Mech.} \textbf{876}, 1108--1128.  

\bibitem{GHB15}
Berry, T, Giannakis, D, Harlim, J. 2015 Nonparametric forecasting of low-dimensional dynamical systems. \textit{Physical Review E}. \textbf{91(3)}, 032915.

\bibitem{BitEtAl17}
Bittracher, A, Koltai, P, Klus, S, Banisch, R, Dellnitz, M, Sch\"utte, C. 2018
Transition manifolds of complex metastable systems: Theory and data-driven computation of effective dynamics.
\textit{Journal of Nonlinear Science}, \textbf{28(2)}, 471--512.

\bibitem{PBLRS20}
Pillaud-Vivien, L, Bach, F, Lelièvre, T, Rudi, A, Stoltz, G. 2020 Statistical estimation of the {P}oincar\'e constant and application to sampling multimodal distributions. In \textit{International Conference on Artificial Intelligence and Statistics}, PMLR, 2753-2763.

\bibitem{qiu_1998}
Qiu XL, Xia KQ. 1998 Spatial structure of the viscous boundary layer in
  turbulent convection.
\textit{Phys. Rev. E} \textbf{58}, 5816--5820.

\bibitem{cioni_1997}
Cioni S, Ciliberto S, Sommeria J. 1997 Strongly turbulent rayleigh–bénard
  convection in mercury: comparison with results at moderate prandtl number.
\textit{J. Fluid Mech.} \textbf{335}, 111–140.

\bibitem{gallet}
Gallet B, Herault J, Laroche C, P\'etr\'elis F, Fauve S.  Reversals of a
large-scale field generated over a turbulent background.
\textit{Geophysical and Astrophysical Fluid Dynamics} \textbf{106} 468 -492.

\bibitem{valencia_2007}
Valencia L, Pallares J, Cuesta I, Grau FX. 2007 Turbulent RayleighB´enard convection of water 
in cubical cavities: a numerical and experimental study.
\textit{International Journal of Heat Mass Transfer} \textbf{50} 3203–3215.

\bibitem{monin_1971} Monin AS, Yaglom AM. 1971  \textit{Statistical Fluid Mechanics I}.  New York, MIT Press.

\bibitem{monin_1975} Monin AS, Yaglom AM. 1975  \textit{Statistical Fluid Mechanics II}.  New York, MIT Press.


\end{thebibliography}

\end{document}